# Capabilities of a 24 -channel slicer for imaging spectroscopy with the Multichannel Subtractive Double Pass (MSDP) on 8-meter class solar spectrographs

*Malherbe, J.-M.*

*Observatoire de Paris - Section de Meudon, LESIA, 92195 Meudon, France*

*December 6, 2023*

### Abstract

Imaging spectroscopy is intended to be coupled with adaptive optics (AO) on large solar telescopes, in order to produce high spatial and temporal resolution measurements of velocities and magnetic fields on a 2D target. We present the theoretical capabilities of a new generation 24-channel MSDP slicer for 8-meter class spectrographs which are common in solar astronomy. The aim is to produce 24-channel spectra-images providing **cubes** of **instantaneous data (x, y, λ)** allowing the study of the plasma dynamics and magnetic fields. We investigate the possibility of doubling the spectral resolution using two interlaced spectra-images, delivering together 48 channels. Two polarimetric methods are also explored providing simultaneous measurements of Stokes combinations with a dual beam; one of them could provide 48 sub-channels (or 96 with wavelength interlaced observations).

### Introduction

The Multichannel Subtractive Double Pass (MSDP) is an imaging spectroscopy technique introduced by Mein (1977). It is based on a slicer which provides line profiles with N sampling points (or N channels) over a 2D field of view (FOV); for that purpose, the MSDP uses a rectangular entrance window instead of a thin slit. The technique was progressively developed and implemented on many telescopes (Mein et al, 2021). The first instrument (N = 7, 9 channels) was incorporated to the 14 m spectrograph of the Meudon Solar Tower (MST). It was mainly running with the Hα line in order to study the dynamics (via the Doppler effect) of chromospheric features (filaments, prominences, active regions). It was followed by the MSDP of the 8 m spectrograph of the Pic du Midi Turret Dome (Mein, 1980) with N = 11 channels. The third instrument was integrated to the 15 m spectrograph of the german Vacuum Tower Telescope (VTT) in Tenerife (Mein, 1991). Meanwhile, polish colleagues introduced the MSDP on the large Bialkow coronagraph to observe prominences (Rompolt, 1994). The next instrument (N = 16 channels) was designed for the 8 m spectrograph of the THEMIS telescope (Mein, 2002). For the first time, the MSDP worked there in polarimetric mode using achromatic waveplates and a calcite beam-splitter. The full Stokes capabilities allowed Mein et al (2009) to investigate the vector magnetic field above an active region.

These instruments used first generation slicers, based on multi-slits in the spectrum and prism beam-shifters. However, this technique does not allow to increase much the spectral resolution (limited to about 80 mÅ) and the number N of channels. For that reason, the second generation of slicers incorporates now micro-mirrors, which allow to reach about 30 mÅ spectral resolution and can deliver more than 50 channels. The first slicer using this technology was installed at MST (N = 18 channels). The Solar Emission Line Dopplerometer, in project for Bialkow and Lomnicky coronagraphs (N = 24 channels) is dedicated to velocity measurements in the hot corona using the forbidden lines of iron (Malherbe et al, 2021). Since a second 24-channel slicer is available, the present paper discusses what could be done with 8-meter class spectrographs, such as those of Pic du Midi, THEMIS, the future European Solar Telescope (EST) or others.

# 1 – Characteristics of the 24-channel, new generation slicer

The 24-channel slicer is based on 24 micro-mirrors for beam-splitting (a manufactured mono-block device) and 24 adjustable shifting mirrors, providing a translation between channels of 4.8 mm in the spectrum. In practice, the useful field in the x-direction is smaller (4.1 mm). The field in the y-direction could be 8 times larger, for example with a typical entrance window of 4.1 x 32.8 mm². These distances (in mm) have a correspondence in arcsec on the Sun which depends on the focal length of the telescope. For instance, in the case of the THEMIS telescope, the 57 m focal length would provide a 17" x 136" FOV; it would be about the same at Pic du Midi. The MSDP spectra-image size in x-direction is 4.8 mm * (24 + 2) = 125 mm (the number 2 corresponds to the separation between the two optical blocks of 12 channels inside the slicer).

The step of the slicer is $\Delta x$ = 0.4 mm. It provides an associated step in wavelength $\Delta x/d$, which depends on the spectrograph dispersion (d in mm/Å) available for the observed line.

We chose below (as an example) six lines of various width from 0.08 Å (thin photospheric lines, as those of FeI) to 1.0 Å (broad chromospheric lines such as Hα), measured at the inflection points (FWIP or Full Width at Inflexion Points, indicated on the graphs below). A spectral field (BW or bandwith) centred on the line core, in the range 0.25 Å to 1.50 Å (according to the selected line) is imposed to observe, at any point of the 2D (x, y) FOV, the line core, inflection points and close wings. It is essential to include inflexion points in the bandpass, because the Stokes V signal is maximum there, according to the weak field theory. Close wings are of interest to compute dopplershifts lower in the atmosphere than in the line core.

For a 8-meter spectrograph with a blazed grating at 63°26', we typically have, with k the interference order:

k $\lambda$ = 226434 in the blaze for a 79 grooves/mm grating, or k $\lambda$ = 59628 in the blaze for 300 grooves/mm.

The lines in the blaze inside the spectral range 4000-8500 Å appear in orders 26-56 for 79 grooves/mm or 7-15 for 300 grooves/mm.

In the blaze, the dispersion is:

d = 0.141 k mm/Å in the blaze for the 79 grooves/mm grating, or

d = 0.537 k mm/Å in the blaze for 300 grooves/mm,

so that the typical dispersion for lines inside the spectral domain 4000-8500 Å decreases from 8 to 4 mm/Å. The spectral resolution $\Delta x/d$ of the slicer is given by table 1 for various spectral lines. The wavelength functions of the slicer for these lines are given in section 2.

As the magnification of the spectrograph is 1.0 (same focal length for the collimator and the camera mirror), the output spectra-images have a width of 125 mm. A lens attached to the CCD or CMOS sensor must reduce the image to 36 mm width or less, according to the detector size (commercial cameras propose usually chips smaller than 24 x 36 mm²).

| $\lambda$ (Å) | FWIP (Å) | d (mm/Å) | $\Delta x/d$ (mÅ) | min BW (Å) |
|---|---|---|---|---|
| 3934 | 0.18 | 8.07 | 50 | 0.50 |
| 5173 | 0.15 | 6.25 | 64 | 0.35 |
| 5896 | 0.20 | 5.36 | 75 | 0.40 |
| 6173 | 0.085 | 5.24 | 76 | 0.25 |
| 6563 | 1.0 | 4.82 | 84 | 1.50 |
| 8542 | 0.4 | 3.68 | 109 | 0.70 |

*Table 1* : spectral resolution $\Delta x/d$ of the slicer for various lines: CaII K 3934, Mg b2 5173, NaD1 5896, FeI 6173, H 6563, CaII 8542. d is the dispersion in the blaze for 8 m focal length. FWIP is the Full Width at Inflexion Points. BW is the chosen bandwidth around the line centre for any point of the rectangular FOV.

## 2 – Spectral line sampling

The figures 1 – 6 below show the wavelength functions of the 24-channel slicer for the spectral lines of table 1 (CaII K 3934, Mg b2 5173, NaD1 5896, FeI 6173, H 6563, CaII 8542). They are given by this formula: $\lambda_n(x) = \lambda_0 + (x/d) + n\,(\Delta x/d)$, for $0 < x < x_m$ ($x_m$ = 3.6 mm is the maximum x-direction FOV); n is the current channel (1 < n < 24), d is the dispersion of the spectrograph (mm/Å); Δx = 0.4 mm is the slicer step (the spectral resolution in Å is Δx/d).

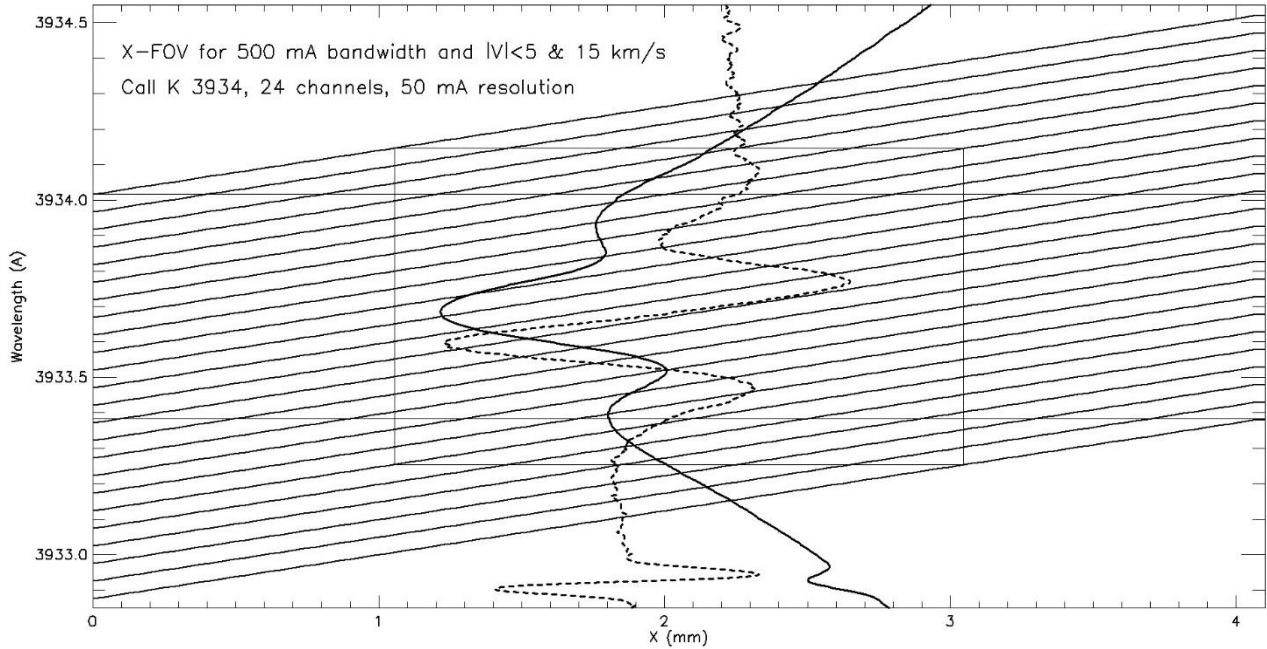

*Figure 1* : wavelength functions of the 24 channels for CaII K 3934. The atlas profile at disk centre is reported together with its derivative to locate inflexion points. Velocities up to 5 km/s can be measured in the full FOV, or less than 15 km/s in the central part. 50 mÅ resolution, 180 mÅ FWIP. Courtesy Paris Observatory.

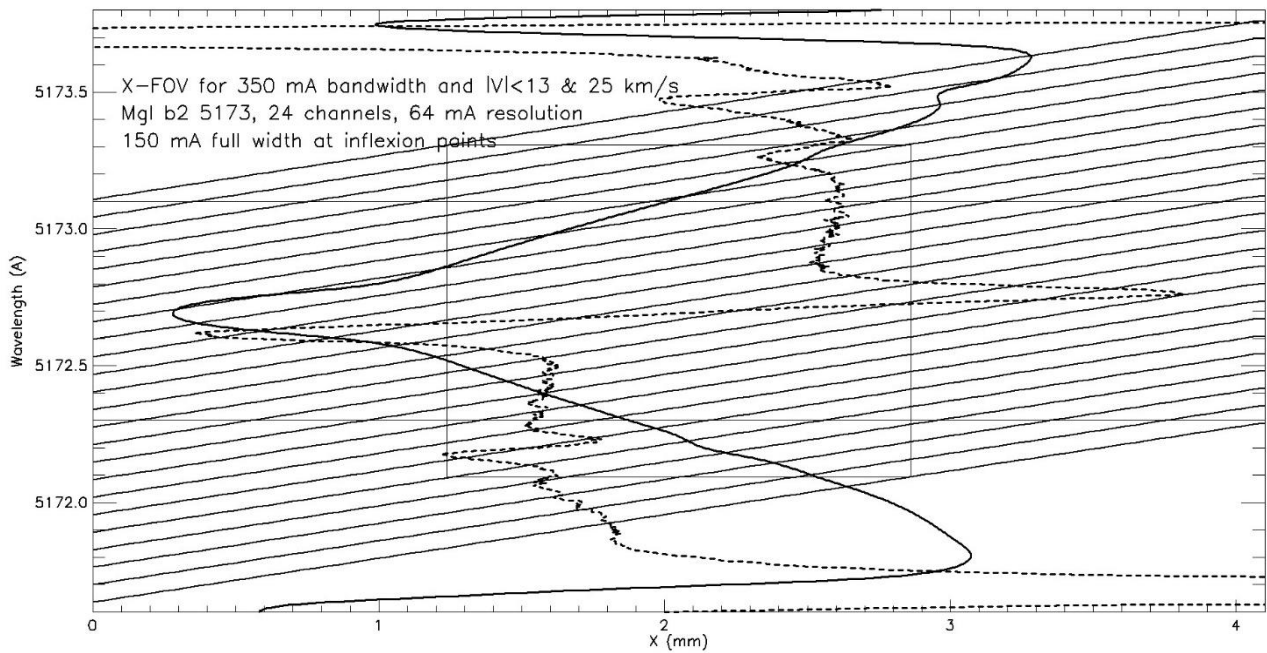

*Figure 2* : wavelength functions of the 24 channels for Mg b2 5173. Velocities up to 13 km/s can be measured in the full FOV, or less than 25 km/s in the central part. 64 mÅ resolution, 400 mÅ FWIP. The atlas profile at disk centre is reported together with its derivative to locate inflexion points. Courtesy Paris Observatory.

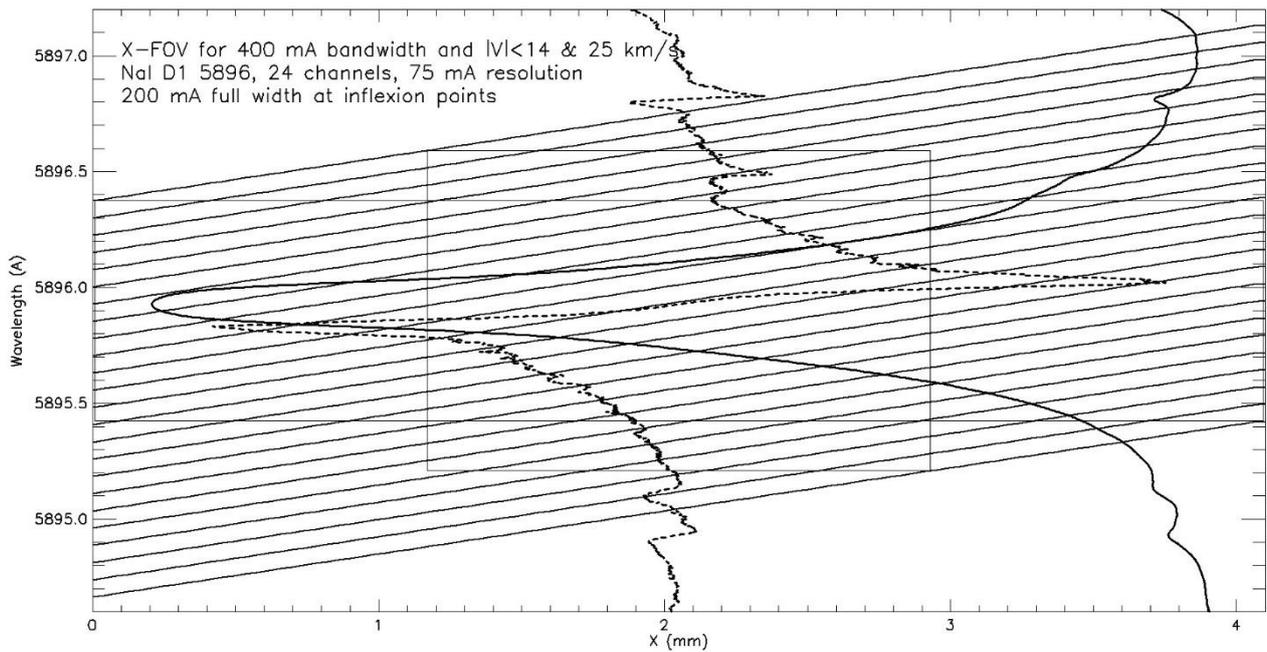

*Figure 3* : *wavelength functions of the 24 channels for Na D1 5896. Velocities up to 14 km/s can be measured in the full FOV, or less than 25 km/s in the central part. 75 mÅ resolution, 200 mÅ FWIP. The atlas profile at disk centre is reported together with its derivative to locate inflexion points. Courtesy Paris Observatory.*

The FeI 6173 line is a thin photospheric line (86 mÅ FWIP). But the 24-channel slicer provides only 76 mÅ spectral resolution, which is not sufficient; however, it is possible to interlace two consecutive observations done in the presence of AO in order to stabilize the FOV and correct the seeing conditions. The second observations could be obtained within one second (or less) after rotating the grating, in order to translate the spectral line of a half step, i.e. 38 mÅ. This method would provide 2 x 24 = 48 interlaced channels with 38 mÅ spectral resolution (figure 4), quite compatible with the width of thin photospheric lines.

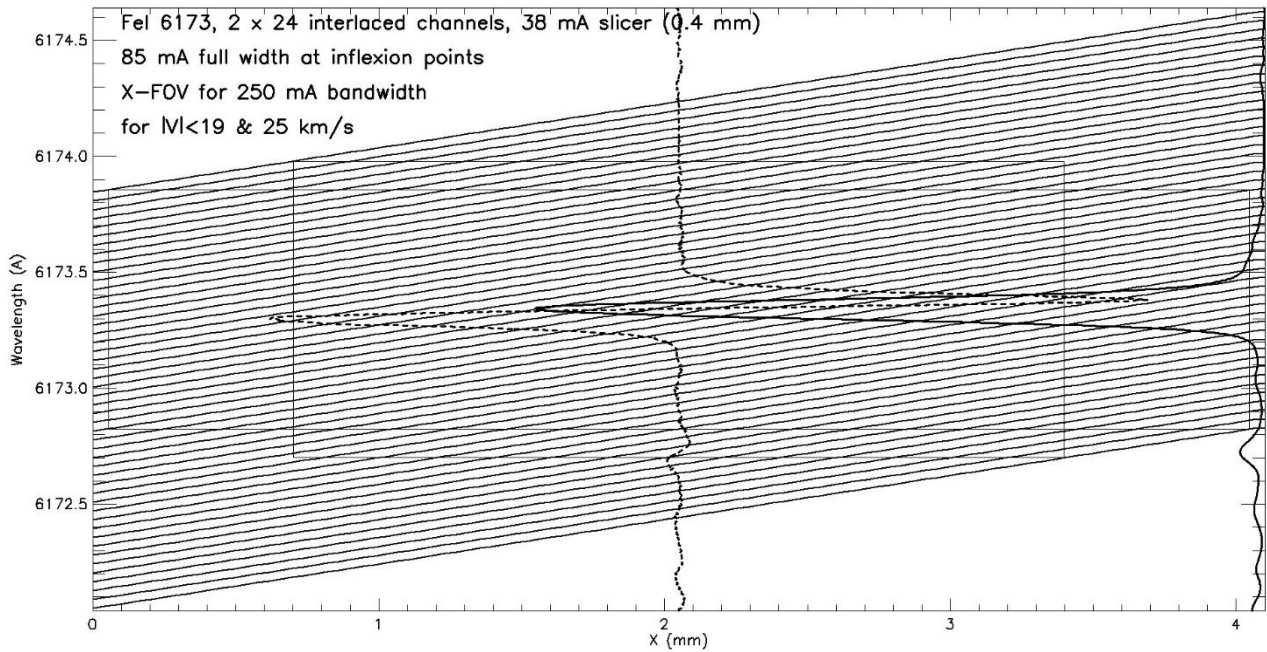

*Figure 4* : *wavelength functions of 2 x 24 interlaced channels for FeI 6173. Velocities up to 19 km/s could be measured in the full FOV, or less than 25 km/s in the central part. 38 mÅ resolution, 85 mÅ FWIP. The atlas profile at disk centre is reported together with its derivative to locate inflexion points. More than two interlaced observations could be possible, if necessary. Courtesy Paris Observatory.*

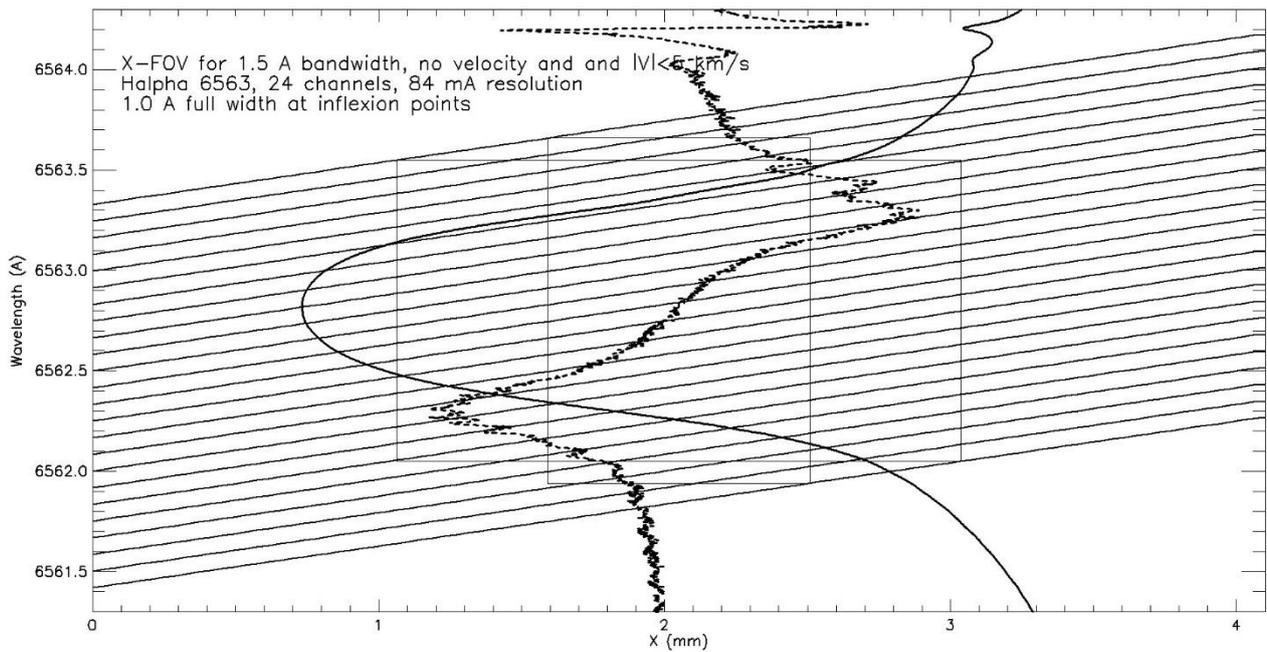

*Figure 5* : wavelength functions of the 24 channels for Hα 6563. Velocities up to 5 km/s can be measured in the central part of the FOV, but there are not enough channels to measure high velocities (the 84 mÅ resolution of the slicer is in fact not necessary for a so broad line of 1.0 Å FWIP). The atlas profile at disk centre is reported together with its derivative to locate inflexion points. Courtesy Paris Observatory.

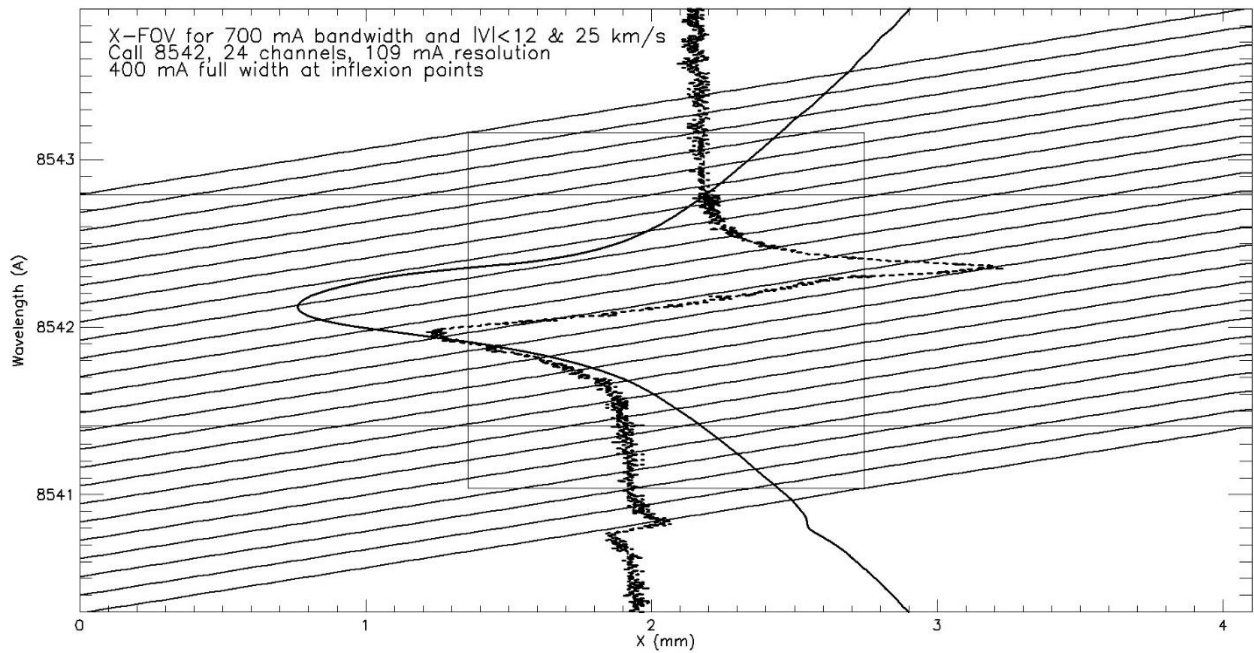

*Figure 6* : wavelength functions of the 24 channels for CaII 8452. Velocities up to 12 km/s can be measured in the full FOV and 25 km/s in the central part of the FOV. 109 mÅ resolution, 400 mÅ FWIP. The atlas profile at disk centre is reported together with its derivative to locate inflexion points. Courtesy Paris Observatory.

Figure 7 shows the six spectral lines and the wavelength sampling points for the central part of the FOV (x = $x_m/2$, black) and the two extreme positions, x = 0 (blue) and x = $x_m$ (red). One sees that spectral lines are well resolved with 24 channels, except FeI 6173 which requires the interlacing of two observations (2 x 24 channels) in order to increase the spectral resolution.

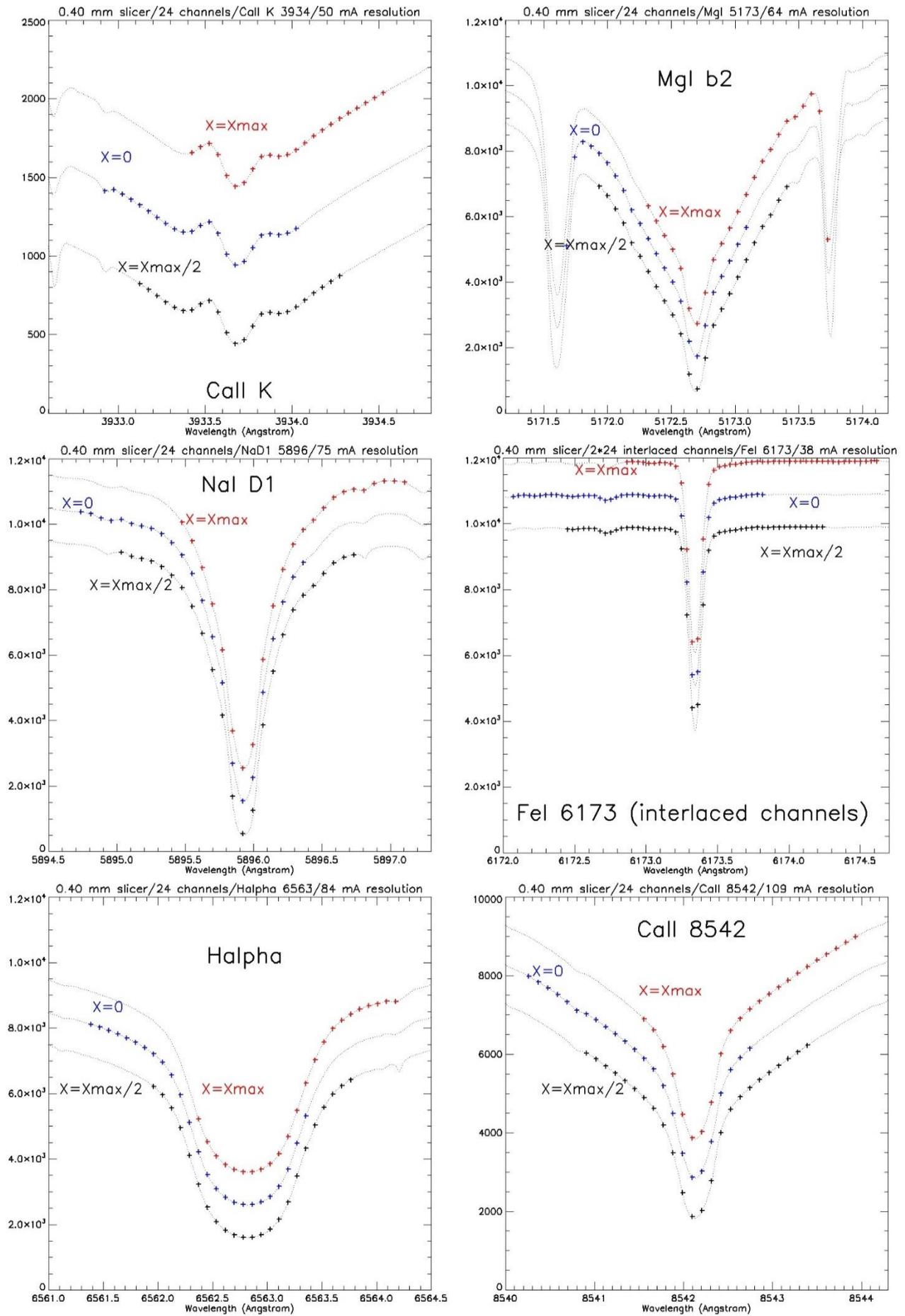

*Figure 7* : spectral line sampling at 3 positions of the FOV ($x = 0$, $x_m/2$, $x_m$). Courtesy Paris Observatory.

### 3 – Simulation of observations with the 24-channel slicer

Meudon spectroheliograph provides observations of the Hα and CaII K lines of the full Sun from spectroscopic scans of the solar surface, with low spatial resolution (2") and moderate spectral resolution (0.155 Å and 0.093 Å respectively). The instrument produces data-cubes (x, y, λ) in which only two coordinates are simultaneous (y, λ). However, available line profiles allow to simulate possible observations of active regions with the MSDP 24-channel slicer. Figure 8 shows the 24 channels over an active region with filaments in the Hα line, while figure 9 displays the 24 channels in CaII K 3934, over the same FOV than Hα.

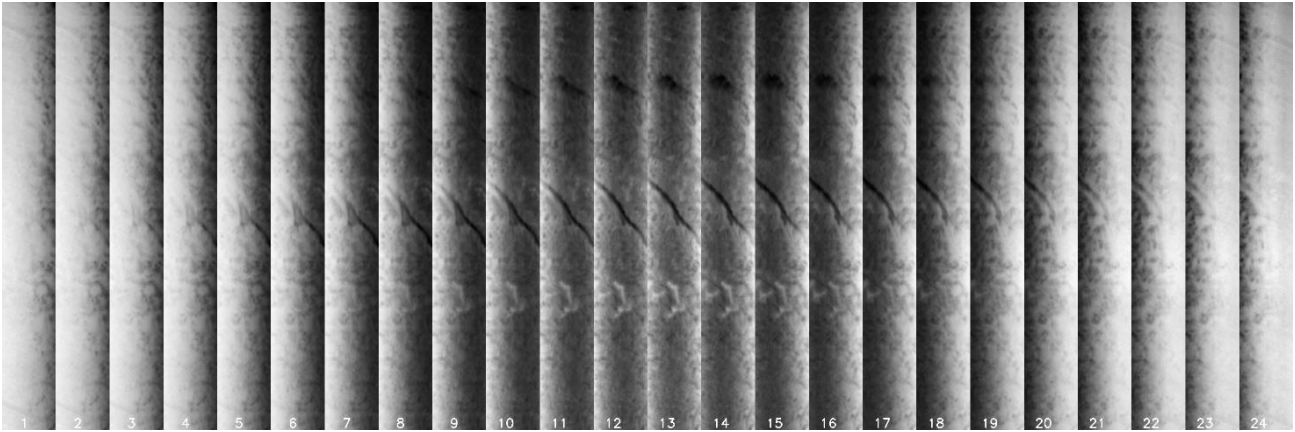

**Figure 8** : simulation of the 24 channels using Hα line profiles from Meudon spectroheliograph. The wavelength functions for each channel are those of figure 5. Courtesy Paris Observatory.

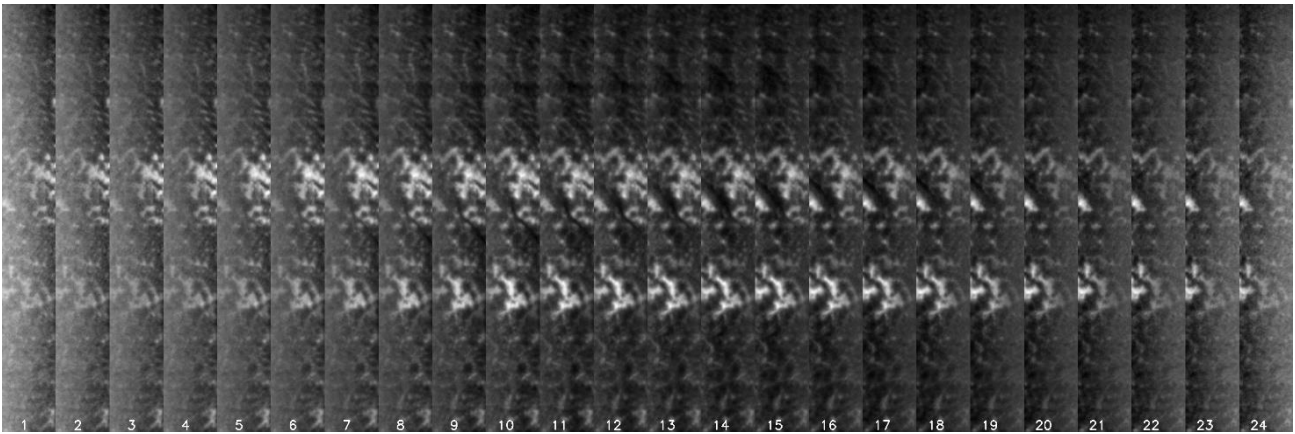

**Figure 9** : simulation of the 24 channels using CaII K line profiles from Meudon spectroheliograph. The wavelength functions for each channel are those of figure 1. Courtesy Paris Observatory.

Concerning the thin FeI 6173 line, the 24-channel slicer with 76 mÅ spectral resolution does not suffice. Figure 10 displays a simulation using the more precise method with 2 x 24 interlaced channels. We assumed that the line profile has a gaussian shape with four parameters (continuum intensity, depression, width, dopplershift). SDO/HMI data (15 September 2023 at 12:00 UT) provided the continuum intensity and the dopplershift. The line width (85 mÅ FWIP) and the central depression (0.63) were assumed to be constant across the FOV. Two successive observations, very close in time and stabilized by AO, are required to interlace even and odd channels. Between both observations, a fast grating rotation of ½ slicer step (38 mÅ) is done. This process delivers two sets of 24 interlaced wavelengths which allow to double the spectral resolution and rebuild the line profiles with 48 sampling points (as shown by figure 7). The wavelength functions for each channel are those of figure 4.

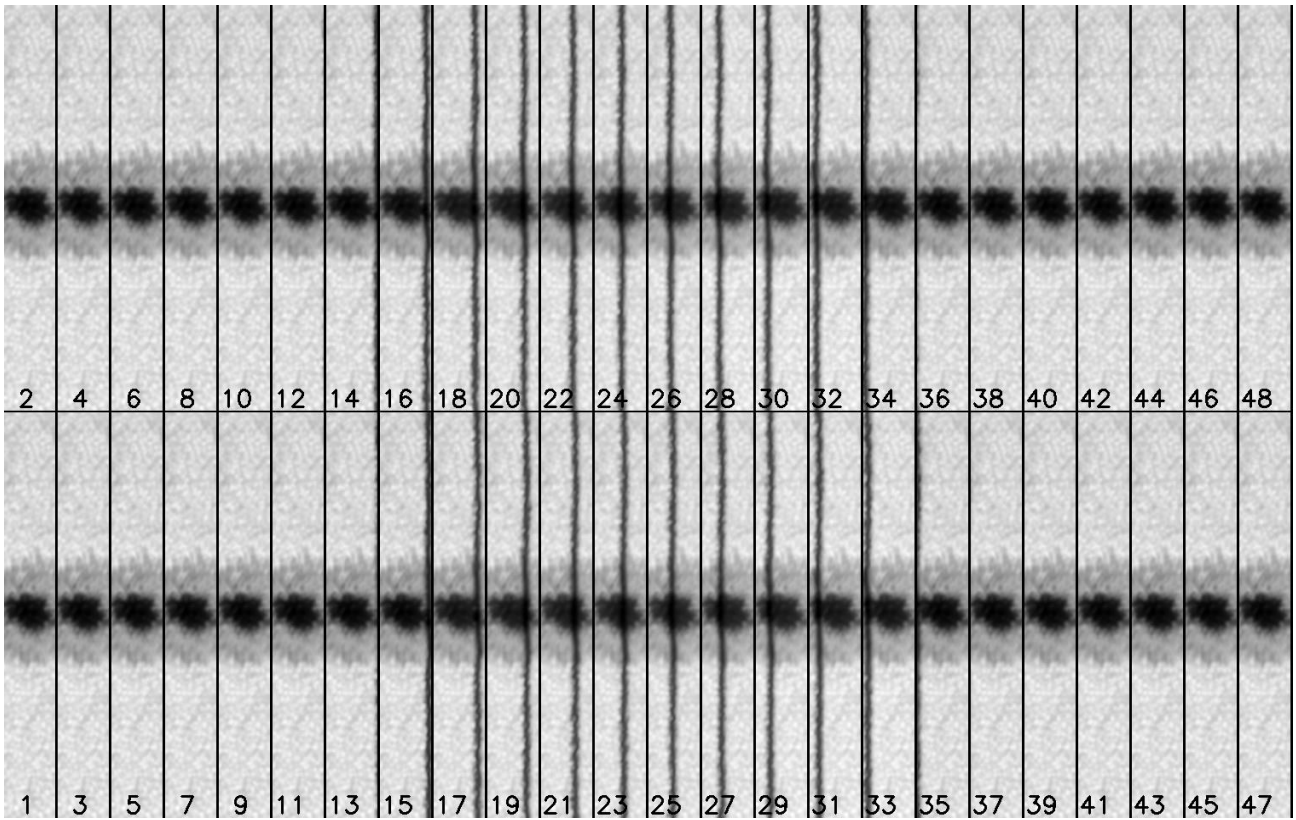

*Figure 10* : simulation of the 2 x 24 interlaced channels based on FeI 6173 SDO/HMI observations. A first series of 24 even channels is followed by a second series of 24 odd channels after fast grating rotation of ½ slicer step (38 mÅ), providing finally 48 sampling points with 38 mÅ resolution. Courtesy Paris Observartory.

**4 – Polarimetric methods using a birefringent beam splitter-shifter**

In order to measure simultaneously Stokes combinations I+S and I-S (where S = Q, U, V in sequence), we propose the dual-beam polarimetric method described by Semel (1980).

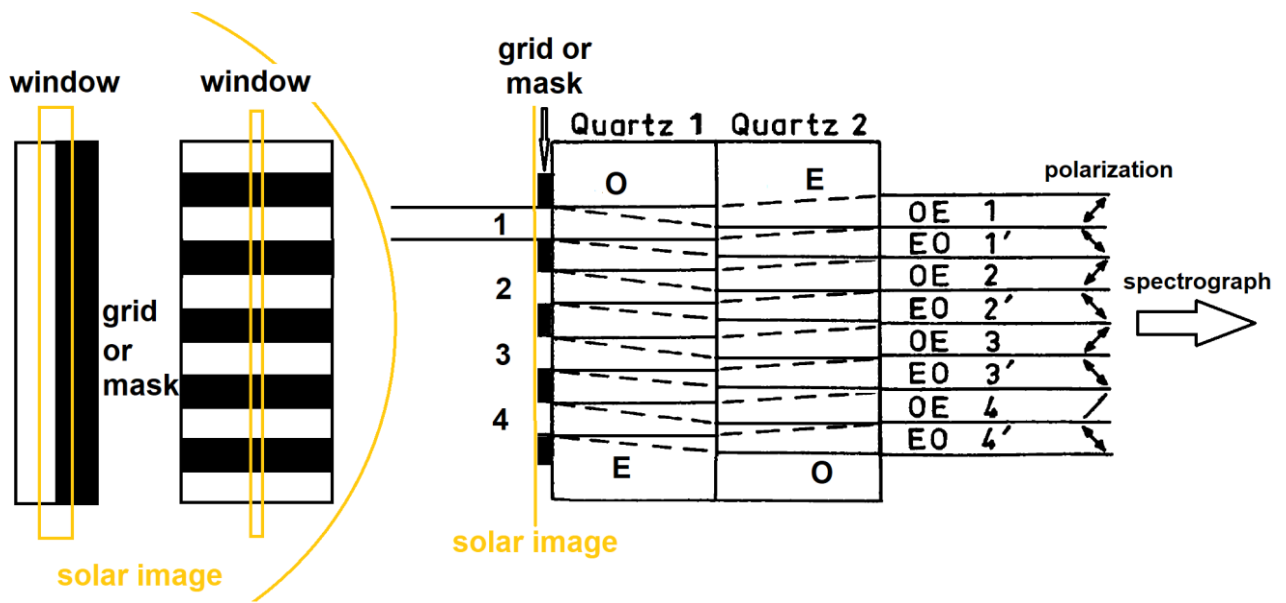

*Figure 11* : the dual-beam polarimetric method using a periodic grid across the entrance window (beam splitting in y-direction), or a mask (beam splitting in x-direction). In both cases, the FOV is reduced by a factor two, so that two consecutive exposures are necessary to recover the full FOV. Courtesy Paris Observartory.

The method consists in hiding half of the FOV by a mask or a grid, in order to form a dual beam with a quartz or calcite polarizing beam splitter-shifter (figure 11). In order to equalize optical paths, two crossed crystals are used; the ordinary ray in the first one corresponds to the extraordinary ray in the second one, and inversely. For instance, beam 1 is separated into two co-spatial beams (1 and 1', respectively I+S and I-S signals) which have orthogonal linear polarizations and are injected into the spectrograph. We experienced either a mask (separation in x-direction) or a grid (separation in y-direction), as emphasized by figure 12.

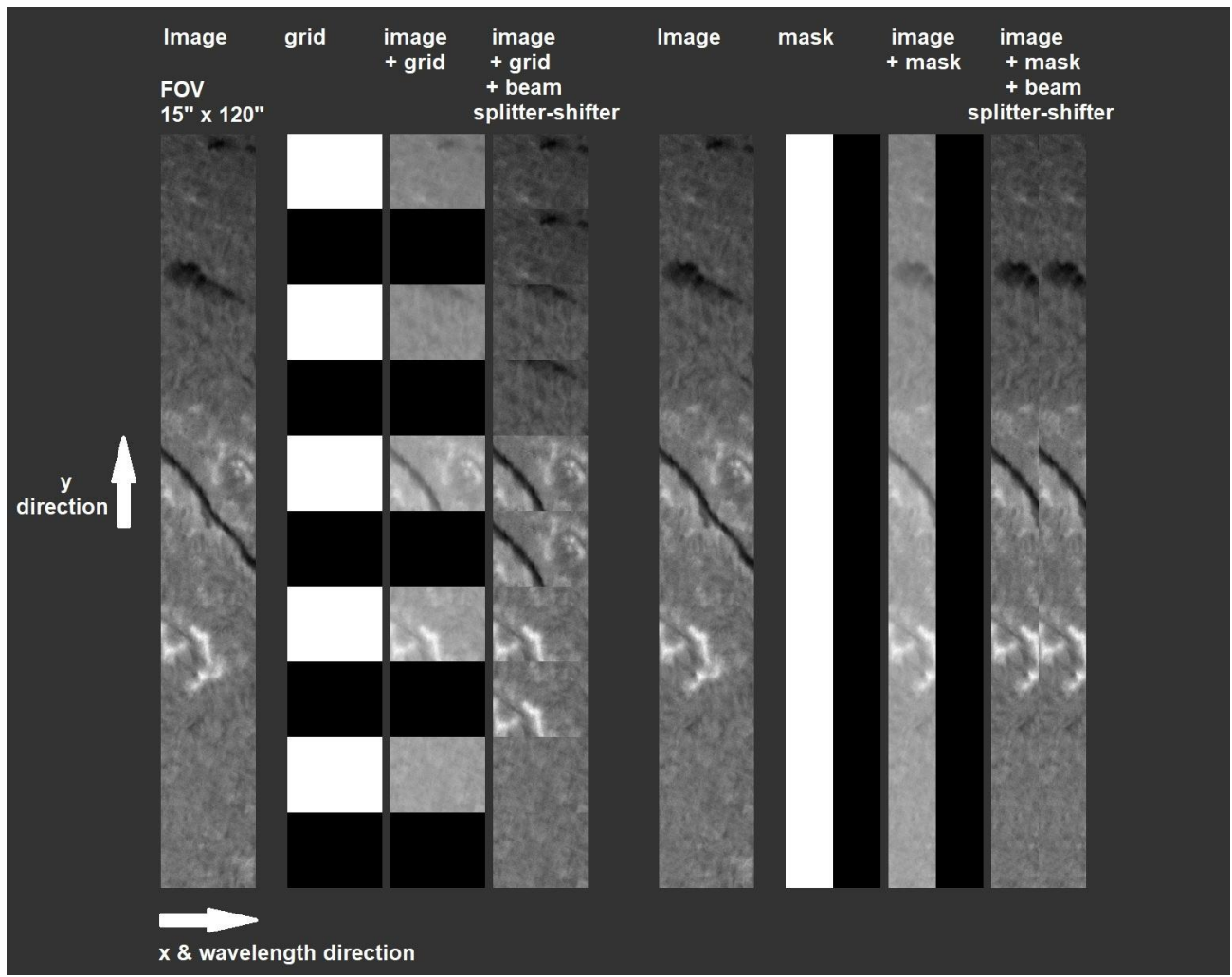

*Figure 12 : the grid method (left) and mask method (right) used with the polarimetric MSDP. With the grid, the separation is the y-direction, while with the mask, it is the x-direction. The figure shows for each case: the observed FOV, the grid (or mask) alone, the observed FOV transmitted by the grid (or mask) implying that the half of the FOV is hidden, and the final image delivered by the birefringent beam splitter-shifter. It is composed of co-spatial sub-images in two simultaneous states of polarization (I+S, I-S, where S = Q, U, V in sequence). Courtesy Paris Observartory.*

In both cases (grid or mask), half of the FOV is lost, to that two consecutive observations are needed to recover the full FOV by moving the telescope (the mask can alternatively be translated in x-direction, or the grid in y-direction). This should be done preferably when AO is available, in order to correct seeing effects.

The grid has the disadvantage of creating many sub-FOVs and the advantage of providing exactly the same wavelength sampling for the output Stokes combinations I+S and I-S.

The mask has the advantage of creating a unique sub-FOV and associated MSDP co-spatial sub-channels, and the disadvantage of providing different wavelength samplings for the output Stokes combinations I+S and I-S. Hence, a wavelength interpolation is required to combine I+S and I-S profiles, in order to derive the profiles of $I(\lambda)$, $S(\lambda)$ and the polarization rate $S/I(\lambda)$. For imaging spectroscopy programs, we prefer this method.

## 5 – Polarimetric method 1 with grid and beam splitting-shifting in the y-direction

We show in this section a simulation of FeI 6173 (figure 13) based on SDO/HMI data (continuum intensity, dopplershifts, magnetic fields) and the interlaced method (wavelength functions of figure 4).

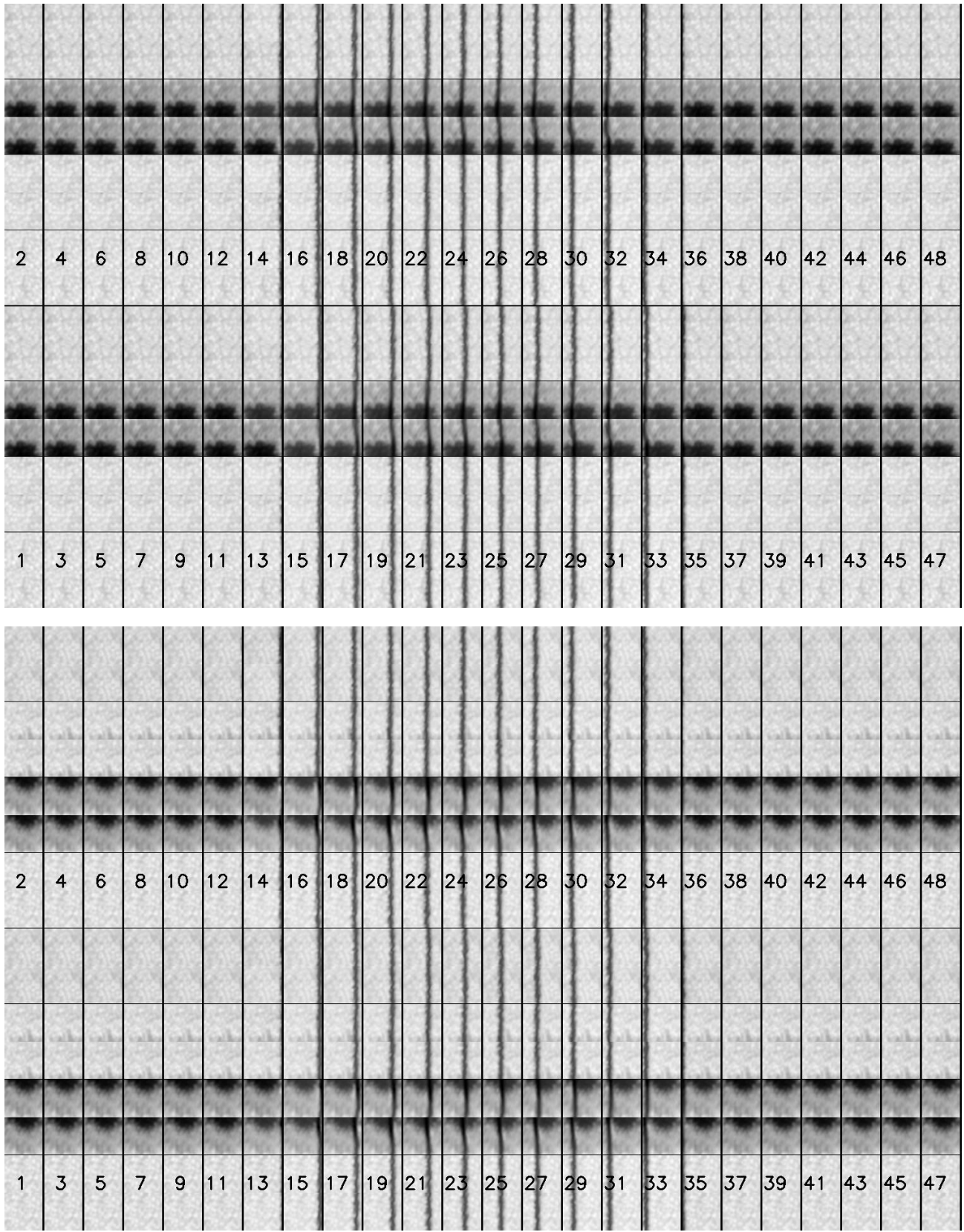

*Figure 13* : simulation of FeI 6173 with 2 x 24 interlaced channels. Even channels are 38 mÅ shifted, so that the final resolution is 38 mÅ. Two observations (top, bottom) are required to cover the full FOV. Courtesy OP.

Figure 13 shows a simulation of two successive observations providing 2 x 24 interlaced channels (76 mÅ spectral resolution for each), which can be combined to produce the final 38 mÅ spectral resolution, and repeated two times to record the full FOV. The polarimeter of figure 11 with the grid and y-direction beam-splitting delivers simultaneous I+V and I-V signals (the Zeeman shift is well visible in the sunspot). Four successive observations are needed (top and bottom parts of figure 13) to restore the full FOV, after telescope or grid displacement. For that simulation, we assumed that line profiles have a gaussian shape (with constant width of 85 mÅ and 0.63 central depression) and used the weak field theory, with continuum intensity, dopplershifts and magnetic fields provided by SDO/HMI. The line profiles of I+V(λ), I-V(λ) and polarization rate V/I(λ) with B// = 500 G are shown in figure 14 for 3 positions in the FOV, x=0, $x_m/2$ and $x_m$.

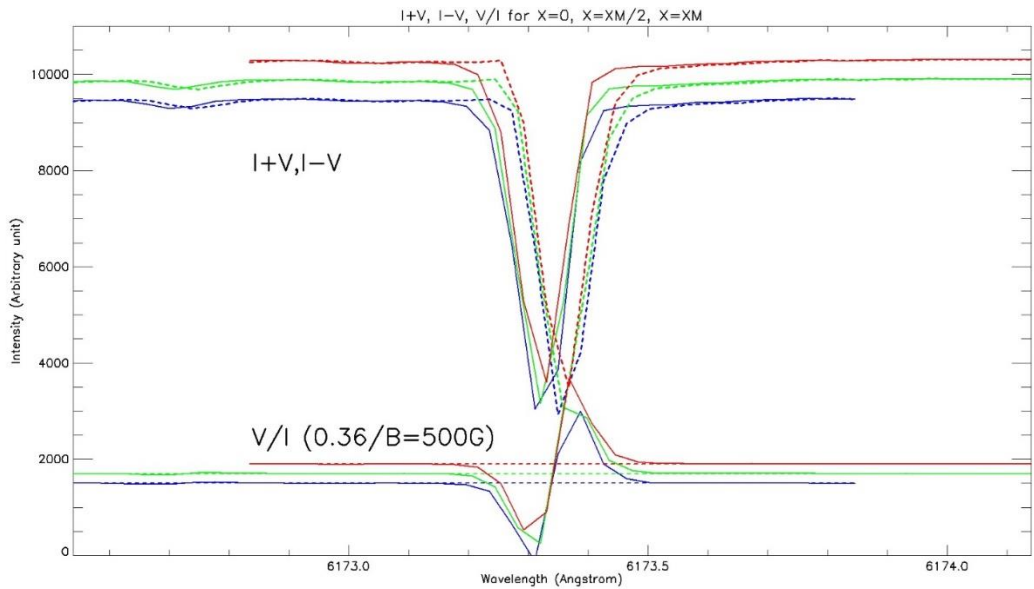

**Figure 14** : simulation of Stokes profiles I+V(λ), I-V(λ) (solid/dashed lines) and polarization rate V/I(λ) with B// = 500 G for FeI 6173 with 2 x 24 interlaced channels and grid (y-direction separation). The spectral resolution is 38 mÅ. The profiles are shown for 3 positions in the FOV: x=0, $x_m/2$ and $x_m$. Courtesy Paris Observatory.

### 6 – Polarimetric method 2 with mask and beam splitting-shifting in the x-direction (sub-channels)

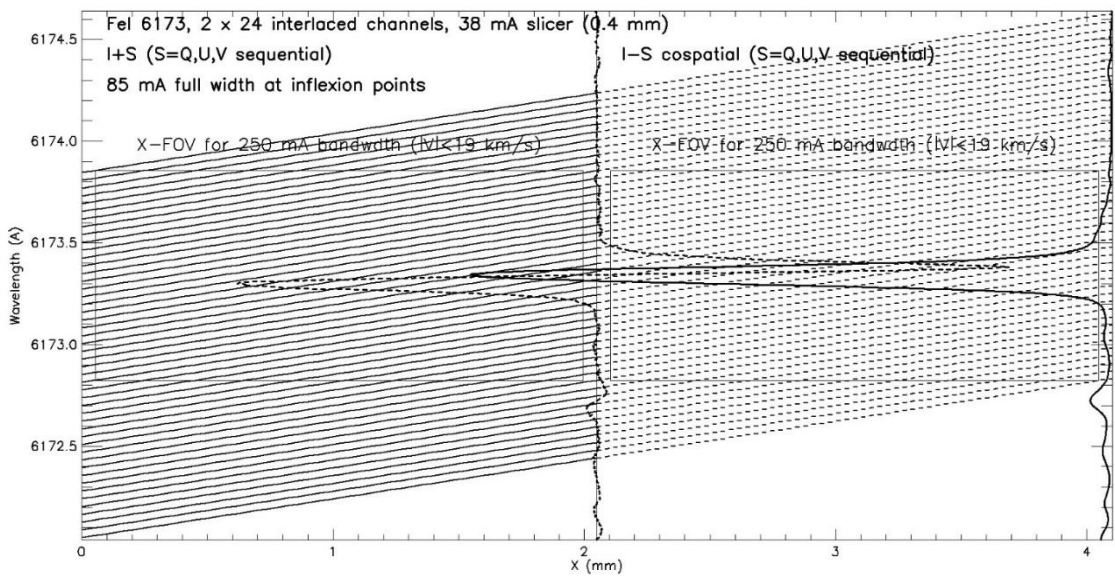

**Figure 15** : wavelength functions of the 2 x 24 interlaced channels for FeI 6173 in the case of half FOV masking for simultaneous Stokes profiles I+S (left part) and I-S (right part) measurements (S = Q, U, V in sequence). Left

*and right regions are strictly co-spatial, but their wavelength sampling is not identical. Velocities up to 19 km/s could be measured everywhere. 38 mÅ resolution, 85 mÅ FWIP. The atlas profile at disk centre is reported together with its derivative to locate inflexion points. Courtesy Paris Observatory.*

With the mask, the calcite of figure 11 is oriented in order to produce an x-direction beam splitting and shifting, so that the interlaced method of 2 x 24 channels delivers spectra-images as those of figure 16.

**Figure 16** : *simulation of FeI 6173 with 2 x 24 = 48 interlaced channels, 2 x 48 = 96 polarimetric sub-channels. The final resolution is 38 mÅ. Two observation (top, bottom) are required to cover the full FOV. Courtesy OP.*

Figure 16 shows a simulation of two successive observations providing 2 x 24 = 48 interlaced channels (even and odd channels, 76 mÅ spectral resolution for each), which can be combined to produce the final 38 mÅ spectral resolution with 48 channels, well adapted to thin lines. The polarimeter of figure 11 with the mask and x-direction beam-splitting delivers simultaneous I+V and I-V signals which appear as 2 x 48 = 96 co-spatial polarimetric sub-channels in figure 16. However, there is a wavelength shift between the sub-channels equal to $x_m/2d$ = 0.39 Å, as revealed by figure 15. As the mask hides half of the FOV, two wavelength interlaced observations are required for each half FOV, so that four successive observations are needed to restore the full FOV (top and bottom parts of figure 16), after telescope or mask displacement. The AO, when available, should be running to avoid seeing fluctuations during the four observations. For that simulation, we assumed also that line profiles have a gaussian shape and used the weak field theory, with continuum intensity, dopplershifts and magnetic fields provided by SDO/HMI. The line profiles of I+V(λ), I-V(λ) do not have the same wavelength sampling so that an interpolation is necessary to recombine them before producing the I(λ) and V(λ) profiles and polarization rate V/I(λ). Examples are shown in figure 17 with B// = 500 G for 3 positions in the half FOV, x=0, $x_m/4$ and $x_m/2$.

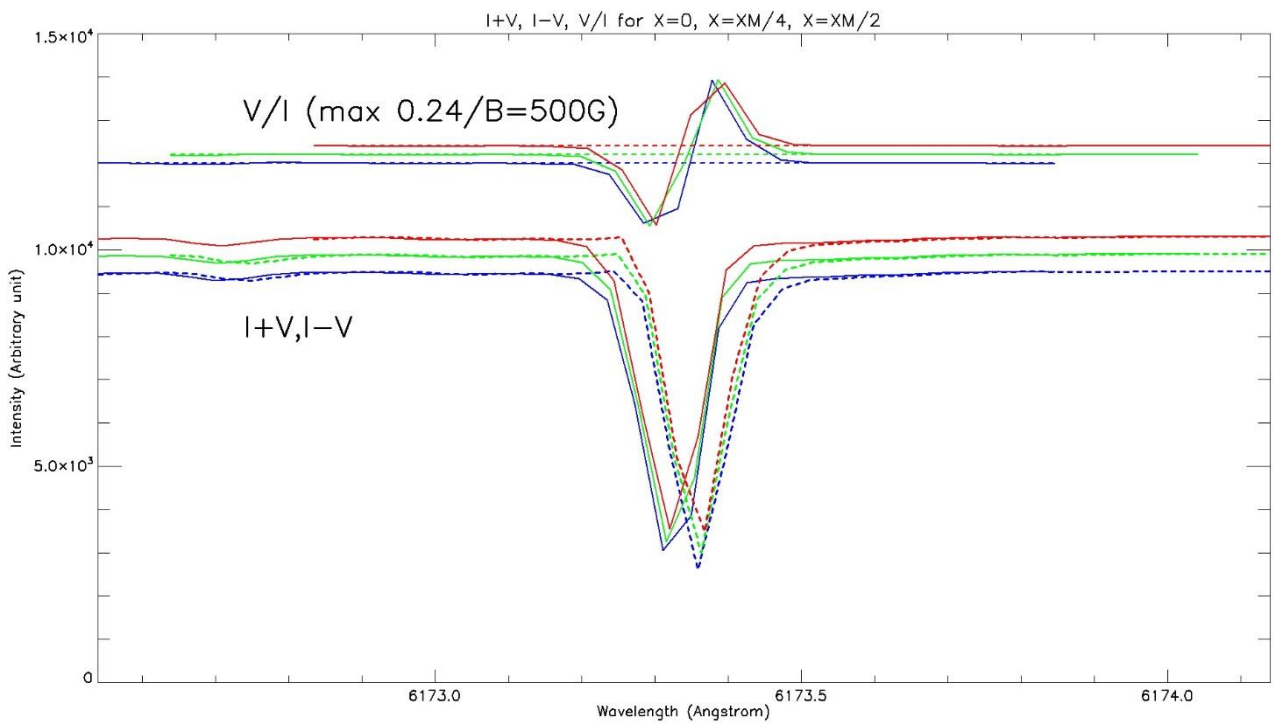

*Figure 17* : simulation of Stokes profiles I+V(λ), I-V(λ) (solid/dashed lines) and polarization rate V/I(λ) with B// = 500 G for FeI 6173 with 2 x 24 interlaced channels and half FOV masking (separation in x-direction). The spectral resolution is 38 mÅ. The profiles are shown for 3 positions in the half FOV: x=0, $x_m/4$ and $x_m/2$. The wavelength sampling of I+V(λ) and I-V(λ) profiles is not the same, they are shifted of $x_m/2d$ = 0.39 Å, so that an interpolation is needed to combine them and compute V/I(λ). Courtesy Paris Observatory.

**Conclusion**

Figure 18 shows a simulation based on either figure 13 or figure 16 for the polarization rate V/I of FeI 6173 using two wavelength interlaced observations (total of 48 wavelength sampling points) and the two polarimetric methods described in previous sections (grid and separation in y-direction, mask and separation in x-direction). Hence, it combines four successive observations to interlace wavelengths and cover the full FOV. Theoretically, both methods are equivalent, but in practice, the mask avoids too many sub-FOVs and creates 96 co-spatial polarimetric sub-channels. The methods decribed in this paper are promising for imaging spectro-polarimetry of both chromospheric lines and thin photospheric lines, such as those of FeI.

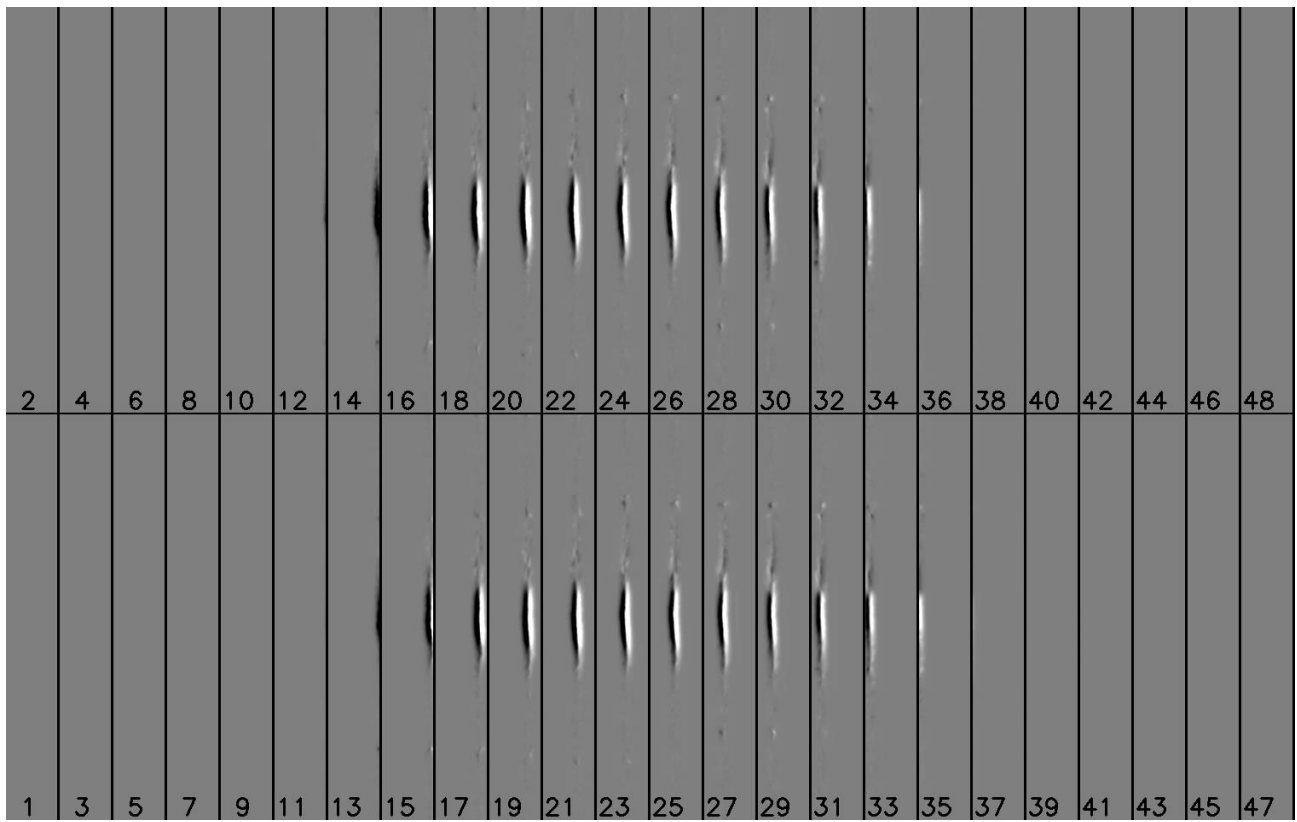

*Figure 18* : simulation of the polarization rate V/I(λ) with B//=500 G of FeI 6173. This image results of the combination of data shown either in *figure 13* or in *figure 16*. Courtesy Paris Observatory.